%% file: SE_CDC_arxiv.tex
\documentclass[10pt]{IEEEtran}
\input{preambleT.tex}
\title{\vspace{1cm}\centering \LARGE \textbf{State Estimation in Water Distribution Networks through a New Successive Linear Approximation}}
\author{Shen Wan$\text{g}^*$, Ahmad F. Tah$\text{a}^*$, Lina Sel$\text{a}^{\ddagger}$, Nikolaos Gatsi$\text{s}^{*}$, and Marcio H. Giacomon$\text{i}^{\dagger}$	
	\thanks{
		$^*$Department of Electrical and Computer Engineering, The University of Texas at San Antonio. $^{\dagger}$Department of Civil and Environmental Engineering, The University of Texas at San Antonio, $^{\ddagger}$ Department of Civil, Architectural and Environmental Engineering, Cockrell School of Engineering, The University of Texas at Austin. Emails: mvy292@my.utsa.edu, \{ahmad.taha, nikolaos.gatsis, marcio.giacomoni\}@utsa.edu, linasela@utexas.edu. This material is based upon work supported by the National Science Foundation under Grant CMMI-DCSD-1728629. }}
\begin{document}

	\newdimen\origiwspc%
	\newdimen\origiwstr%
	\origiwspc=\fontdimen2\font
	\origiwstr=\fontdimen3\font
	
	\fontdimen2\font=0.63ex
	
	\maketitle	
	\begin{abstract}
		State estimation (SE) of water distribution networks (WDNs) is difficult to solve due to nonlinearity/nonconvexity of water flow models, uncertainties from parameters and demands, lack of redundancy of measurements, and inaccurate flow and pressure measurements. This paper proposes a new, scalable successive linear approximation to solve the SE problem in WDNs. The approach amounts to solving either a sequence of linear or quadratic programs---depending on the operators' objectives. The proposed successive linear approximation offers a seamless way of dealing with valve/pump model nonconvexities, is different than a first order Taylor series linearization, and can be incorporated into with robust uncertainty modeling.
		Two simple test-cases are adopted to illustrate the effectiveness of proposed approach using head measurements at select nodes.
	\end{abstract}
	
	%
	
	\section{Introduction and Paper Contributions}~\label{sec:literature}
	Water distribution networks (WDNs) are designed to deliver water to various residential and business consumers with sufficient pressure and flow~\cite{zhao2016optimization}. The calculation of flows and heads/pressures in WDNs can be obtained by the principles of \textit{conservation of mass} and \textit{energy}. The former implies the continuity of flow at nodes, and the latter states that energy difference stored in a component equals the energy increases minus energy losses, such as, frictional and minor losses~\cite{puig2017real}. 
	
	The challenging part of monitoring WDNs is that pipes are usually buried underground and are inaccessible~\cite{tshehla2017state}. Hence, it is impossible to monitor the flow in all pipes and the head at all nodes even with modern supervisory control and data acquisition (SCADA) systems, let alone enable continuous monitoring of WDNs, which is limited in practice due to high investment, operations and maintenance costs~\cite{aisopou2012pipe}. 
	
	A practical approach to gain a network-wide observability, while addressing the aforementioned limitations, is to use state estimation (SE), which can determine the unknown variables of a system based on a set of local measurements and a hydraulic network model~\cite{cheng2014real}. Usually, the set of measurements consists of heads at key nodes and the flows through key links. However, the SE problem is difficult due to uncertainty from pipe roughness coefficients, demands, and measurement errors~\cite{Vrachimis2018RealtimeHI}. One way to reduce uncertainty is by introducing redundancy of observations, which significantly improves the performance of the SE procedure. The degree of redundancy is achieved by combining actual measurements (e.g., heads and flows) with the pseudo-measurements (e.g., demands); however, due to limited measurement availability, the application of SE algorithms to WDNs is an ongoing research~\cite{tshehla2017state}. 
	
	In WDNs, the SE problem is predominantly cast as an inverse problem to determine unknown system conditions with an objective, e.g., \textit{weighted least-squares} (WLS), to minimize the mismatch between measurements and hydraulic model estimations~\cite{bargiela1984line}. The authors in~\cite{powell1999state} discuss a way to obtain the solution from over-determined measurements. The study~\cite{davidson2006adjusting} produces solutions that are consistent with available SCADA data by adjusting estimated demands based on WLS method. The authors in~\cite{barkdoll2003effect} use Monte Carlo simulation (MCS) to evaluate the effect of variable demands on pressure and water quality, and their work is extended by~\cite{pasha2005analysis}. In order to overcome the computational time of MCS, a new approximate method for uncertainty analysis  is proposed in~\cite{kang2009approximate}. 
	
	The authors in~\cite{fusco2017state} propose a SE in the presence of control devices with switching behavior, such as pressure reducing valves after a minor modification of existing WLS solvers. An approach combining regression-trees with genetic algorithms to fit demands to the observations was proposed in~\cite{preis2011}.  In~\cite{Vrachimis2018RealtimeHI}, the authors solve the real-time SE problem using interval linearization of the nonlinear flow equations and successively tightening the interval bounds. In summary, the SE problem results in nonlinear and nonconvex system of equations, which exhibit serious scalability issues when applied to realistic WDNs. 
	\begin{table*}[t]
		\def\arraystretch{1.3}
		\caption{Hydraulic models of pipes and pumps and their converted models (time index $k$ is ignored for each variable for simplicity).}
		\centering
		\makegapedcells
		\setcellgapes{0.2pt}
		\begin{tabular}{c|c|c}
			\hline
			& {\textit{Pipes}}  & \textit{Pumps} \\ \hline
			{\textit{Original Hydraulic Model}}
			&  \parbox{7cm}{
				\vspace{-0.7em}
				\begin{align}~\label{equ:head-flow-pipe}
				\setlength{\abovedisplayskip}{3pt} 
				\setlength{\belowdisplayskip}{3pt}
				\Delta h_{ij}^\mathrm{P}  = h_{i} - h_{j} = R_{ij} {q_{ij}}|q_{ij}|^{\mu-1}
				\end{align}
				\vspace{-1em}
			}
			& 
			\parbox{6cm}{
				\vspace{-0.7em}
				\begin{align} \label{equ:head-flow-pump}
				\hspace{-10pt} \Delta h_{ij}^\mathrm{\mathrm{M}} = h_{i} - h_{j} = -{s_{ij}^2}(h_0 - r  (q_{ij} s_{ij}^{-1})^\nu )
				\end{align}
				\vspace{-1em}
			} 
			\\ 
			\hline
			
			{\textit{GP Form}} 
			& 
			\parbox{7cm}{
				\begin{align}~\label{equ:head-loss-pipe-exp}
				{\hat{h}_{i}} {\hat{h}_{j}^{-1}} [\widehat{C}^{\mathrm{P}}_{ij}]^{-1} {\hat{q}_{ij}}^{-1} = 1
				\end{align}
			} 
			&\parbox{6cm}{
				\vspace{-0.7em}
				\begin{align} \label{equ:head-flow-pump-exp}
				\hspace{-12pt}{\hat{h}_{i}} {\hat{h}_{j}^{-1}}[\widehat{C}_1^{\mathrm{M}}]^{-1}[{\hat{q}_{ij}}]^{-\widehat{C}_2^{\mathrm{M}}}  = 1
				\end{align}
				\vspace{-1em}
			} \\ \hline
			{\textit{Linear Form}} 
			& \parbox{7cm}{
				\vspace{-0.7em}
				\begin{align} \label{equ:head-flow-pipe-linear}
				\Delta h_{ij}^\mathrm{P}  = h_{i} - h_{j} = C_{ij}^{\mathrm{P}}+ {q_{ij}}
				\end{align}
				\vspace{-1em}
			} 
			&\parbox{6cm}{
				\vspace{-0.7em}
				\begin{align} \label{equ:head-flow-pump-linear}
				\hspace{-10pt}\Delta h_{ij}^\mathrm{\mathrm{M}} = h_{i} - h_{j} = C_1^{\mathrm{M}} + C_2^{\mathrm{M}} q_{ij}
				\end{align}
				\vspace{-1em}
			}
			\\ \hline \hline
		\end{tabular}
		\label{tab:models}%
		\vspace{-0em}
	\end{table*}

	This paper proposes a new, scalable successive linear approximation to solve the SE problem in WDNs. The approach amounts to solving either a sequence of linear or quadratic programs (LP/QP), depending on the operator's objectives. The proposed successive linear approximation offers a seamless way of dealing with valve/pump model nonconvexities, is different than a first order Taylor series linearization, and can be easily incorporated into uncertainty modeling. The paper's contributions can be summarized as follows:
	
	\noindent $\bullet$ The classical, highly nonlinear and nonconvex state estimation problem is converted into successive convex LP/QP problems via \textit{geometric programming} (GP) approximation~\cite{duffin:gp}. 
	
	\noindent $\bullet$ A new optimization technique is introduced to solve the nonconvex SE problem through a tractable computational algorithm for a general WDN topology. The proposed research builds on our recent work on pump control of WDNs using GP~\cite{wang2019geometric}, but offers a different approach through LP formulation, in comparison with our prior work.
	
	The paper organization is given next. Section~\ref{sec:Modeling} describes SE formulation. In Section~\ref{sec:Conversion}, the proposed optimization-based SE technique is introduced, conversion of nonconvex SE into LP/QP is given, and two test-cases are used to illustrate the effectiveness of our approach in Section~\ref{sec:Test}. Finally, Section~\ref{sec:Future} presents the paper's limitations and future research directions.
	\section{Modeling and State Estimation of WDNs}~\label{sec:Modeling}
	WDN is modeled by a directed graph $(\mathcal{V},\mathcal{E})$.  Set $\mathcal{V}$ defines nodes and is partitioned as $\mathcal{V} = \mathcal{J} \bigcup \mathcal{T} \bigcup \mathcal{R}$ where $\mathcal{J}$, $\mathcal{T}$, and $\mathcal{R}$ stand for the collection of junctions, tanks, and reservoirs. Let $\mathcal{E} \subseteq \mathcal{V} \times \mathcal{V}$ be the set of links, and define the partition $\mathcal{E} = \mathcal{P} \bigcup \mathcal{M} \bigcup \mathcal{W}$, where $\mathcal{P}$, $\mathcal{M}$, and $\mathcal{W}$ stand for the collection of pipes, pumps, and valves. For the $i^\mathrm{th}$ node, set $\mathcal{N}_i$ collects its neighboring nodes and is partitioned as $\mathcal{N}_i = \mathcal{N}_i^\mathrm{in} \bigcup \mathcal{N}_i^\mathrm{out}$, where $\mathcal{N}_i^\mathrm{in}$ and $\mathcal{N}_i^\mathrm{out}$ stand for the collection of inflow and outflow nodes. According to the principles of \textit{conservation of mass} and \textit{energy}, we present the modeling in WDNs next.
	\setlength{\abovedisplayskip}{2pt} 
	\setlength{\belowdisplayskip}{2pt}
	\subsection{Modeling WDNs}~\label{sec:Model_iass}
	In this section, we introduce the modeling of WDNs.
	\subsubsection{Tanks and Reservoirs}	
	The water hydraulic dynamics in the $i^\mathrm{th}$ tank can be expressed by a discrete-time difference equation~\cite{wang2019geometric}
	\begin{align}~\label{equ:tankhead}
	\hspace{-14pt}h_{i}^{\mathrm{TK}}(k+1)\hspace{-2pt} =\hspace{-2pt} h_{i}^{\mathrm{TK}}(k) \hspace{-2pt}+\hspace{-2pt} \frac{\Delta t}{A_i^{\mathrm{TK}}}\hspace{-3pt}\left(\hspace{-1pt}\sum_{j \in \mathcal{N}_i^\mathrm{in}}\hspace{-3pt}q_{ji}(k)\hspace{-2pt}-\hspace{-7pt}\sum_{j \in \mathcal{N}_i^\mathrm{out}} \hspace{-3pt}q_{ij}(k)\hspace{-3pt}\right)\hspace{-3pt}.
	\end{align}
	where  $h_i^{\mathrm{TK}}$, $A_i^{\mathrm{TK}}$ respectively stand for the head, cross-sectional area  of the $i^\mathrm{th}$ tank, and $\Delta t$ is sampling time;   $q_{ji}(k),\;j \in \mathcal{N}_i^\mathrm{in} $ is inflow, while $q_{ij}(k),\;j \in \mathcal{N}_i^\mathrm{out} $ is outflow of the $j^\mathrm{th}$ neighbor. We assume that reservoirs have infinite water supply and the head of the $i^\mathrm{th}$ reservoir is fixed as $h_i^{\mathrm{R}_\mathrm{set}}$~\cite{zamzam2018optimal,rossman2000epanet} which is perfectly accurate. This also can be viewed as an operational constraint~\eqref{equ:tank-reservoir}.
	
	\subsubsection{Junctions and Pipes}
	Junctions are the points where water flow merges or splits. The expression of mass conservation of the $i^\mathrm{th}$ junction  can be written as
	\begin{align}~\label{equ:nodes}
	\sum_{j \in \mathcal{N}_i^\mathrm{in}} q_{ji} (k)- \sum_{j \in \mathcal{N}_i^\mathrm{out}} q_{ij} (k)= d_i(k),
	\end{align}
	where $d_i(k)$ stands for end-user demand that is extracted from node $i$ at time $k$.
	
	The major head loss of a pipe from node $i$ to $j$ is due to friction and is determined by~\eqref{equ:head-flow-pipe} from Tab.~\ref{tab:models}, where $R$ is the pipe resistance coefficient and $\mu$ is the constant flow exponent, both are determined by the corresponding formula, Hazen-Williams, Darcy-Weisbach, or Chezy-Manning. The approach we proposed considers any of the three formulae~\cite{rossman2000epanet}. Minor head losses are not considered in this paper, but can be easily modeled through equivalent pipe length.
	
	\subsubsection{Head Gain in Pumps} 	A head increase/gain can be generated by a pump between suction node $i$ and delivery node $j$. Generally, the head gain can be expressed as~\eqref{equ:head-flow-pump}, where $h_0$, $r$, and $\nu$ are the pump curve coefficients; $q_{ij}$ is the flow through a pump; $s_{ij} \in [0,s_{ij}^{\mathrm{max}} ]$ is the relative speed of the pump, we assume that the speed is fixed and can be expressed as $s_{ij} = s_{ij}^{\mathrm{max}} =1$. Notice that head gain $h_{ij}^M$ is always negative, and can be viewed as an operational constraint~\eqref{equ:headgainlimit}.
	
	For all the operational limitations of head at each junction and flow though each pipe, we list them as~\eqref{equ:headLimit}. Hence, the compact constraints are
	\begin{subequations} ~\label{equ:constraints} 
		\begin{align}
		\hspace{-2.5cm}{\mathrm{Constraints:}}\; \; \;  \; \;  \; \;  \; \;  \; \;   h_i^{\mathrm{R}}(k) &= h_i^{\mathrm{R}_\mathrm{set}}~\label{equ:tank-reservoir}\\
		h_{ij}^{M} (k)&\leq 0~\label{equ:headgainlimit}\\
		h_{i}^{\mathrm{min}} \leq  h_{i} (k)\leq h_{j}^{\mathrm{max}},&\	q_{ij}^{\mathrm{min}} \leq  q_{ij}(k) \leq q_{ij}^{\mathrm{max}}.  ~\label{equ:headLimit}
		\end{align}
	\end{subequations}
	\vspace{-1.8em}
	\subsection{State Estimation Formulation}\label{sec:SE-WDN}
	Classical state estimation problems are typically presented as
	\begin{equation}~\label{equ:classical}
	{\m y}= \m g(\m \xi) + \m \epsilon,
	\end{equation}
	where ${\m \xi}$ is the unknown variable,  vector ${\m y}$ includes all measured quantities, the $\m g(\m \xi)$ is the model of system including nonlinear functions, and  the $\m \epsilon$ represents error between true model and measured values via sensors. As we mentioned in Section~\ref{sec:literature}, it is impossible to measure flows and pressures in the entire WDN, except for key locations. Hence, $\m \xi$ can be  a vector collecting all unknown variables and defined as
	\begin{equation}~\label{equ:compactedVar}
	\hspace{-0.3cm}\m \xi(k)  \triangleq \Bigl \lbrace \m h^{\mathrm{J}}(k), \m h^{\mathrm{R}}(k), \m h^{\mathrm{TK}}(k), \m q^{\mathrm{P}}(k), \m q^{\mathrm{M}}(k)   \Bigr \rbrace.
	\end{equation}
	where $\m h^{\mathrm{J}}$, $\m h^{\mathrm{R}}$, and $\m h^{\mathrm{TK}}$ collects the heads at junctions, reservoirs, and tanks; $\m q^{\mathrm{P}}$ and $\m q^{\mathrm{M}}$ collects the flow through pipes and pumps. The ${\m y}$ can be treated as the vector collecting several measured key heads in the scenario of WDNs (sensors are assumed available to measure head). The overall WDN-SE problem can now be written as
	\begin{subequations} \label{equ:se}  
		\begin{align}
		\hspace{-1em}\textbf{WDN-SE}:\;	\min_{\substack{\m \xi(k)}}  f(\m \epsilon)& = \textstyle  \sum_{k=1}^{T}{ \m \epsilon^\top(k)} \m W(k) \m \epsilon(k) \notag \\
		\mathrm{s.t.}\; 	\m h^{\mathrm{TK}}(k+1) &= \m h^{\mathrm{TK}}(k) + \m E^{\mathrm{TK}} \m q^{\mathrm{P}}(k)  \label{equ:tankdynamic} \\
		\m d (k) &= \m E_q \begin{bmatrix} 
		\m q^{\mathrm{P}}(k)  \\
		\m q^{\mathrm{M}}(k) 
		\end{bmatrix}    \label{equ:conservationofmass} \\
		&	\mathrm{Constraints}~\eqref{equ:constraints} & \label{equ:consts} 
		\end{align}
	\end{subequations}
	where $k$ is time-index; $T$ is time-horizon; $\m \xi(k)$ for all $k=1,\ldots, T$ is the optimization variable that includes unmeasured heads $\m h$ and flows $\m q$; $n_e $ represents the number of measurements; $\m \epsilon(k) \in \mathbb{R}^{n_e \times 1}$ is the error to be minimized; $f(\m \epsilon)$ is a WLS objective function and 
	\begin{align*}
	\m \epsilon(k) = \underbrace{\m E_{h} \begin{bmatrix}
		{\m \Delta \m h^{\mathrm{P}}(k) } =  \m \Phi^{\mathrm{P}}(\m q^{\mathrm{P}}(k) )    \\ 
		{\m \Delta \m h^{\mathrm{M}}}(k) = \m \Phi^{\mathrm{M}}(\m q^{\mathrm{M}}(k) ) 
		\end{bmatrix}}_{\text{\normalsize $\m g (\m \xi)$}} - \underbrace{ \m \Delta  \widetilde{\m h}(k)}_{\text{\normalsize $\m y$}},
	\end{align*}
	where $ \m \Phi^{\mathrm{P}}(\m \cdot)$  and $ \m \Phi^{\mathrm{M}}(\m \cdot)$ collect the nonlinear head loss~\eqref{equ:head-flow-pipe} of all pipes and the nonlinear head gain~\eqref{equ:head-flow-pump} of all pumps. The residual $\m \epsilon$ is reminiscent of the model in~\eqref{equ:classical} and captures the error between the true model and differences $\m \Delta  \widetilde{\m h}(k)$ between head measurements, while matrix $\m E_{h}$ is related to the position of sensors. 
	
	The weight matrix at time $k$ is given by a diagonal matrix $\m W(k)$: smaller diagonal elements in $\m W(k)$ imply more accurate measurements. In practice, sensors are usually fixed in key nodes, and accuracy of sensors can also be assumed as fixed. Hence, $\m W(k)$  is assumed to be a constant matrix. The objective function is thus designed to minimize the weighted error, and we refer to $\m W$ as \textit{accuracy matrix}. The constraints in \textbf{WDN-SE} are discussed next.
	
	Equation~\eqref{equ:tankdynamic} collects the dynamics~\eqref{equ:tankhead} of all tanks in the network, and $\m E^{\mathrm{TK}}$ is formed by the coefficients of flows in~\eqref{equ:tankhead}. In fact, this constraint can be added and removed according to the situation; e.g., this constraint can be removed when performing single period analysis. Equation~\eqref{equ:conservationofmass} collects~\eqref{equ:nodes} at all nodes, where matrix $\m E_{q}$ is defined by WDNs topology, and vector $\m d(k)$ collects water demands at all junctions. We do not consider demand uncertainty, and thus $\m d(k)$ is assumed known and perfectly accurate. Constraint~\eqref{equ:consts} includes the linear constraints presented in Section~\ref{sec:Model_iass}; i.e., the heads of reservoirs are usually fixed and equal to their elevation. Hence, we assume that the measurement of $h^\mathrm{R}$ is very accurate. 

	Notice that any head difference can be expressed as the linear combination of nonlinear models of pipes and pumps using $\m E_{h}$. This key observation is thoroughly illustrated in Fig.~\ref{fig:8nodenetwork} in Section~\ref{sec:Test}. We note the following about the \textbf{WDN-SE} problem.

	\noindent $\bullet$ Two scenarios exist in SE problem in WDNs~\cite{powell1999state}: \textit{Scenario 1} is described by having sufficient measurements, e.g., all states can be determined if heads at tanks and reservoirs are known, see the blue line in Fig.~\ref{fig:8nodenetwork}. In fact, the SE in this scenario has equal number of variables and equations. \textit{Scenario 2} refers to the case with over-determined equations, e.g., additional sets of head are measured at several key nodes besides the head at tanks and reservoirs, see the red line in Fig.~\ref{fig:8nodenetwork}. Numerical tests are given in Section~\ref{sec:Test} for both scenarios. 
	
	\noindent $\bullet$  \textbf{WDN-SE} Problem~\eqref{equ:se} is nonconvex due to the nonlinearity and nonconvexity of head loss/gain models $\m \Phi^{\mathrm{P}}(\m \cdot)$ and $\m \Phi^{\mathrm{M}}(\m \cdot)$ of pipes and pumps---the nonconvexity shows up in the objective function, rather than the constraints.  The only optimization variable in \textbf{WDN-SE} is $\m \xi(k)$, and other variables such as $\m \Delta \m h^{\mathrm{P}}(k)$ are expressions of vectors included in~$\m \xi$~\eqref{equ:compactedVar}. Finally, certain variables in $\m \xi(k)$, notably the measured heads at reservoirs and tanks, are considered to be known.
	
	\noindent $\bullet$  While \textbf{WDN-SE} pertains to SE given a batch of measurements for $k=1, \ldots, T$ and then reconstructs the estimates $\m \xi(k)$ for that time-period, a simple windowing algorithm can yield near real-time state estimates. 
	\section{New Linear Approximation of \textbf{WDN-SE}}~\label{sec:Conversion}
	In this section, we propose a new method inspired by geometric programming to convert the nonlinear head loss model~\eqref{equ:head-flow-pipe} and head gain model~\eqref{equ:head-flow-pump} into GP constraints which can also be rewritten as linear constraints. 
	A basic introduction to GP is given at first, and then a new optimization technique related with GP is proposed for ensuing sections.
	\vspace{-0.20cm}
	\subsection{Geometric Program and A New Optimization Technique}~\label{sec:GPmodeling}
	A geometric program~\cite{boyd2007tutorial} is a type of optimization problem can be expressed as
	\begin{align}~\label{equ:GP-standard}
	\textit{GP:}\;\; \min_{\m x >0 } \hspace{15pt} &f_0(\m x) \notag  \\
	\mathrm{s.t.}\hspace{15pt}& f_i(\m x) \leq 1, i = 1,\cdots, m \\
	& g_i(\m x) = 1, i = 1,\cdots, p, \notag 
	\end{align}
	where $\m x$ is an entry-wise positive optimization variable, $f_i(\m x)$ are posynomial functions and $g_i(\m x)$ are monomials.
	
	One main requirement of the GP formulation is  the positiveness of the decision variables, which limits some decision variables and physical constraints in our setting, e.g., flows in pipes and headloss equation. To overcome this modeling limitation we are inspired by linear programming (LP) techniques. In the simplex method\cite{lustig1994interior}, for example, the free variables are split into a positive and negative part, both being nonnegative. In our case, we introduce  an exponential function $f(x) = b^x$ to convert a nonpositive variable to a positive one, since $f(x)$ is always positive. Using this technique, we can convert some problems with negative feasible regions into a new problem with a positive feasible region, and then solve it by using modern optimization solvers. This technique has been successfully applied to solve the control of WDNs in our recent work~\cite{wang2019geometric}. The SE problem here is similar to the control problem of WDNs; however, in the current paper, we  convert the SE problem~\eqref{equ:se} into an LP or QP problem instead of a GP, which provides more elegant---and computationally more efficient---solutions. 
	\subsection{ Conversion of  Energy Balance Equations}~\label{sec:modelconversion}
	Based on the newly introduced optimization technique in Section~\ref{sec:GPmodeling}, we first convert the nonlinear hydraulic model of WDNs into its GP form and then into its LP form. 
	Here,  we convert the head  at the $i^\mathrm{th}$ node $h_i$ and the flow $q_{ij}$ into positive values ${\hat{h}_i}$ and ${\hat{q}_{ij}}$ through exponential functions, ${\hat{h}_i} \triangleq  {b}^{h_i}, \; {\hat{q}_{ij}} \triangleq {b}^{q_{ij}}$, where $b=1+\delta$ is a constant base and $\delta$ is a small positive number. Notice that we only need to convert the  $ \m \Phi^{\mathrm{P}}(\m \cdot)$  and $ \m \Phi^{\mathrm{M}}(\m \cdot)$ into linear form because others are already linear. Tab.~\ref{tab:models} show detailed conversions of all physical models that are all discussed in the following sections.
	
	Next, we convert the pipe model, and let ${\hat{h}_{ij}^{\mathrm{P}}}$ be the GP form of head loss of a pipe, which is obtained  by exponentiating both sides of~\eqref{equ:head-flow-pipe} as follows
	\begin{align}
	{\hat{h}_{i}} {\hat{h}_{j}^{-1}} &= b^{q_{ij} \left(R_{ij}{|q_{ij}|}^{\mu-1} - 1\right)}\ {\hat{q}_{ij}}= \widehat{C}_{ij}^{\mathrm{P}}(q_{ij})\ {\hat{q}_{ij}},\notag
	\end{align}
	where  $\widehat{C}_{ij}^{\mathrm{P}}(q_{ij})=b^{q_{ij} \left(R_{ij}{|q_{ij}|}^{\mu-1} - 1\right)}$ is a function of $q_{ij}$.  Hence, the head loss constraint for each pipe can be written as a monomial equality constraint, which is expressed as~\eqref{equ:head-loss-pipe-exp}, if a an estimate of $\widehat{C}_{ij}^{\mathrm{P}}(q_{ij})$ is known. In order to make it linear, we can execute the $\log$ function on both sides of~\eqref{equ:head-loss-pipe-exp} and obtain~\eqref{equ:head-flow-pipe-linear} where $\widehat{C}_{ij}^{\mathrm{P}}(q_{ij})$ turns into $C_{ij}^{\mathrm{P}}(q_{ij}) = {q_{ij} (R_{ij}{|q_{ij}|}^{\mu-1} - 1)}$.
	
	We note that the expression above is linear with respect to $C_{ij}^{\mathrm{P}}(q_{ij})$ if ${q_{ij}}$ is known, hence we develop a method to find ${q_{ij}}$  by sequentially updating ${q_{ij}}$ and $C_{ij}^{\mathrm{P}}(q_{ij})$. The technique is introduced here. At first, we can make an initial guess denoted by $\langle{q_{ij}}\rangle_0$ for the $0^\mathrm{th}$ iteration ($\langle{C_{ij}^{\mathrm{P}}}\rangle_0$ can be obtained if $\langle{q_{ij}}\rangle_0$ is known), thus, for the $n^\mathrm{th}$ iteration, the corresponding values are denoted by $\langle{q_{ij}}\rangle_n$ and $\langle{C_{ij}^{\mathrm{P}}}\rangle_n$. If the flow rates are close to each other between two successive iterations, we can approximate $\langle{C_{ij}^{\mathrm{P}}}\rangle_n$ using  $\langle{C_{ij}^{\mathrm{P}}}\rangle_{n-1}$, that is
	$\langle{C_{ij}^{\mathrm{P}}}\rangle_n \approx \langle{C_{ij}^{\mathrm{P}}}\rangle_{n-1}.$
	Then, for each iteration $n$, we have  
	\begin{align*} 
	\langle{C_{ij}^{\mathrm{P}}}\rangle_n = {\langle{q_{ij}}\rangle_{n-1} \left(R {|\langle{q_{ij}}\rangle_{n-1}|}^{\mu-1} - 1\right)},
	\end{align*}
	and it can be approximated by a constant given the flow value $\langle{q_{ij}}\rangle_{n-1}$ from the previous iteration. 
	
	Similarly, the new variables ${\hat{q}_{ij}} = b^{q_{ij}}$ and ${\hat{s}_{ij}} = b^{s_{ij}}$ for $(i,j) \in \mathcal{M}$ are introduced for pumps.  Let ${\hat{h}_{ij}^\mathrm{M}} $ be the GP form of head increase of a pump
	\begin{align*} 
	{\hat{h}_{i}} {\hat{h}_{j}^{-1}}  =b^{-{s_{ij}^2} h_0}\ (b^{q_{ij}})^{ r q_{ij}^{\nu-1} s_{ij}^{2-\nu}} 
	={\widehat{C}_1^{\mathrm{M}}} \ ({\hat{q}_{ij}})^{\widehat{C}_2^{\mathrm{M}}},
	\end{align*}
	where $\widehat{C}_1^{\mathrm{M}} = b^{-{s_{ij}^2} h_0}$ and $\widehat{C}_2^{\mathrm{M}} = r q_{ij}^{\nu-1} s_{ij}^{2-\nu}$. Hence, the approximating equation for the pump head increase becomes the monomial equality constraint~\eqref{equ:head-flow-pump-exp} in Tab.~\ref{tab:models}. After executing $\log$ function on both sides of~\eqref{equ:head-flow-pump-exp}, the equation~\eqref{equ:head-flow-pump-linear} can be obtained, which is a linear constraint. And at the same time, the parameters $\widehat{C}_1^{\mathrm{M}}$ and $\widehat{C}_2^{\mathrm{M}}$ become ${C}_1^{\mathrm{M}}$ and ${C}_2^{\mathrm{M}}$, that is
	\begin{align*}
	{C}_1^{\mathrm{M}} &=-{s_{ij}^2} h_0,\,{C}_2^{\mathrm{M}} = r q_{ij}^{\nu-1} s_{ij}^{2-\nu}.
	\end{align*}
	Parameters $C_1^{\mathrm{M}}$ are fixed, while $C_2^{\mathrm{M}}$ follow a similar iterative process as $C_{ij}^{\mathrm{P}}$. That is, starting with an initial guess for the flow rates and relative speeds, the constraints are approximated at every iteration via constraints abiding by the linear form, as listed in Tab.~\ref{tab:models}. This process continues until a termination criterion is met. The details are further discussed in Algorithm~\ref{alg:gp-alg}.
	\subsection{LP/QP Formulation of SE}
	After the conversion of pipe and pump model constraints, we can express the converted problem as
	\begin{subequations} \label{equ:se-LPQP}  
		\begin{eqnarray}
		\hspace{-1.8em} \textbf{LP/QP-SE}: \;\;  \; \;& \hspace{-10em}\min_{\substack{\m \xi(k)}}   \hspace{3em} f(\m \epsilon)   \\
		& \hspace{-8em}\mathrm{s.t.}  \hspace{3em}
		\eqref{equ:tankdynamic}-\eqref{equ:consts}\\
		&	{\m \Delta \m h^{\mathrm{P}}(k)} = \m q^{\mathrm{P}} (k) + \m C^{\mathrm{P}}(k) \label{equ:nolinearPipe-QP} \\    
		&	{\m \Delta \m h^{\mathrm{M}}(k)} = \m C_1^{\mathrm{M}}(k)  + \m C_2^{\mathrm{M}}(k)   \m q^{\mathrm{M}}(k)  \label{equ:nolinearPump-QP}
		\end{eqnarray}
	\end{subequations}
	where constraints and variables remain the same as in~\eqref{equ:se} except that constraints~\eqref{equ:nolinearPipe-QP} and~\eqref{equ:nolinearPump-QP} are now linear and viewed as constraints. The parameters $\m C^{\mathrm{P}}$ is a $\mathbb{R}^{n_p \times 1}$ vector collecting the $C_{ij}^{\mathrm{P}}$ for each pipe. Similarly, the $\m C_1^{\mathrm{M}}$ and $ \m C_2^{\mathrm{M}}$ are a $\mathbb{R}^{n_m \times n_m}$ diagonal matrices collecting $C_1^{\mathrm{M}}$ and $C_2^{\mathrm{M}}$ for each pump.
	\setlength{\textfloatsep}{0.0cm}
	\begin{algorithm}[t]
		\small	\DontPrintSemicolon
		\KwIn{WDN topology, $\langle{{\m \xi}}\rangle_{0}$, demand  $\{\m d(k)\}_{k=1}^{T}$, measurements of head  $\m   \widetilde{\m h}$, the accuracy matrix $\m W$}
		\KwOut{The estimated state value $\{\m \xi_{\mathrm{SE}}(k)\}_{k=1}^{T}$ }
		Set ${\m \xi_{\mathrm{save}}} := \langle{\m \xi}\rangle_{0}$, $n=1$, $\mathrm{step} = 4$, $a = 3$\;
		\While {  $\mathrm{error} \geq \mathrm{threshold}$ \textbf{OR} $n\leq \mathrm{maxIter}$ }{
			Obtain $\langle{C_{ij}^{\mathrm{P}}}\rangle_n$, $\langle{C_{1}^{\mathrm{M}}}\rangle_n$, and $\langle{C_{2}^{\mathrm{M}}}\rangle_n$ from $\langle{\m {\xi}}\rangle_{n-1}$\;
			Generate constraints and form it as \textsc{\textbf{LP/QP-SE}}~\eqref{equ:se-LPQP}\;
			Solve~\eqref{equ:se-LPQP} and obtain $\langle{\m {\xi}}\rangle_{n}$  \;
			\If{$\mathrm{mod}(n,\mathrm{step})=0$}{
				$\m \Delta \m \xi = \langle{\m {\xi}}\rangle_{n}-\langle{\m {\xi}}\rangle_{n-2}$\;
				$\langle{\m {\xi}}\rangle_{n}= \langle{\m {\xi}}\rangle_{n} + a  \m \Delta \m \xi $
			}
			Calculate $\mathrm{error} := \mathrm{norm}(\langle{\m {\xi}}\rangle_{n}-{\m \xi}_{\mathrm{save}})$\;
			Update ${\m \xi_{\mathrm{save}}}=\langle{\m {\xi}}\rangle_{n}$ and $n=n+1$\;
		}
		Set ${{\m \xi}_{\mathrm{SE}}}=\langle{\m {\xi}}\rangle_{n}$
		\caption{Successive approximation of \textbf{WDN-SE}.}
		\label{alg:gp-alg}
	\end{algorithm}
	\setlength{\floatsep}{0.0cm}
	
	We note the following: \textit{(i)} \textbf{LP/QP-SE}~\eqref{equ:se-LPQP} is only an approximation of \textbf{WDN-SE}~\eqref{equ:se} at a specific point (the flow through pipes and pumps $q_{ij}$), in other words, the nonconvex \textbf{WDN-SE} can be approximated by successive convex \textbf{LP/QP-SE}. \textit{(ii)} The converted model is linear but it is not the equivalent to the first order Taylor series linearization. We present the geometric meaning of the conversion we applied via a concrete example in Section~\ref{sec:sufficient}. 
	\textit{(iii)} \textbf{LP/QP-SE} can be expressed as either an LP or QP depending on the objective function. When $f(\m \epsilon)$ is modeled through the absolute weighted error, i.e., $\textstyle \sum_{k=1}^{T} \textstyle \sum_{i=1}^{n_e} w_i(k) |\epsilon_i(k)|$, the problem can be written as an LP. When it is based on WLS, then it becomes a standard QP. 
	
	\subsection{Iterative Algorithm}~\label{sec:Algorithm}
	In order to solve all the unknown variables, our algorithm needs to know the basic information at first, e.g., the topology of tested network to form the matrices $\m E_{q}$, demand ${\m d}$, measurements of head $\m \widetilde{\m h}$, and the accuracy matrix $\m W$. 
	
	
	Notice that all variables are collected in $\m {\xi}$ by~\eqref{equ:compactedVar} and the notation $\langle{{\m \xi}}\rangle_{n}$ in Algorithm~\ref{alg:gp-alg} stands for the $n^\mathrm{th}$ iteration value  $\m {{\xi}}$.  For the $0^{\mathrm{th}}$ iteration, we initialize all flow $\langle{\m {q}^{\mathrm{P}}}\rangle_{0}$ and $\langle{\m {q}^{\mathrm{M}}}\rangle_{0}$ in $\langle{\m {{\xi}}}\rangle_{0}$ with historical average flows. In fact, this algorithm still works by initializing all the flow in $\langle{\m {{\xi}}}\rangle_{0}$ with random number. However, the convergence is relatively slow. All initial statuses of pumps, tanks, and reservoirs are initialized with the value set in ``.inp" source file which is a standard input file used by EPANET, e.g., the initial status (open or close) and speed (if open) of pumps, and the initial head value $\langle{\m {h}^{\mathrm{R}}}\rangle_{0}$ and $\langle{\m {h}^{\mathrm{TK}}}\rangle_{0}$ of tanks and reservoirs. The parameters $\langle{C_{ij}^{\mathrm{P}}}\rangle_1$, $\langle{C_{1}^{\mathrm{M}}}\rangle_1$, and $\langle{C_{2}^{\mathrm{M}}}\rangle_1$ are then calculated by initialized values according to~Section~\ref{sec:modelconversion}, and all constraints are automatically generated for different WDNs topologies. 
	
	After solving~\eqref{equ:se-LPQP} and obtaining the current solution $\langle{\m {{\xi}}}\rangle_{n}$, and defining the iteration error as the Euclidean distance between two consecutive iterations, we set $\langle{\m {{\xi}}}\rangle_{n}$ as the saved value for error calculation in next iteration by assigning ${{\m \xi}_{\mathrm{save}}}=\langle{\m {{\xi}}}\rangle_{n}$. The iteration continues until the error is less than a predefined threshold ($\mathrm{threshold}$) or a maximum number of iterations ($\mathrm{maxIter}$) is reached, and the final solution is ${{\m \xi}_{\mathrm{SE}}}$.
	\begin{figure}[t]
		\centering
		\subfigure[3-node network.]{%
			\centering
			\includegraphics[width=0.55\linewidth]{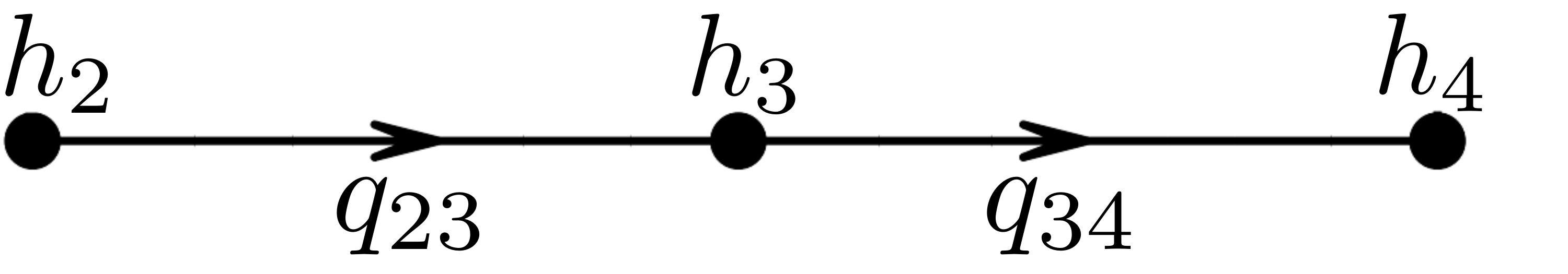}
			\label{fig:3nodePipes}
		} 
		\\ 
		\subfigure[Visualization of equation~\eqref{equ:se-classical}  of 3-node network.]{%
			\includegraphics[width=0.68\linewidth]{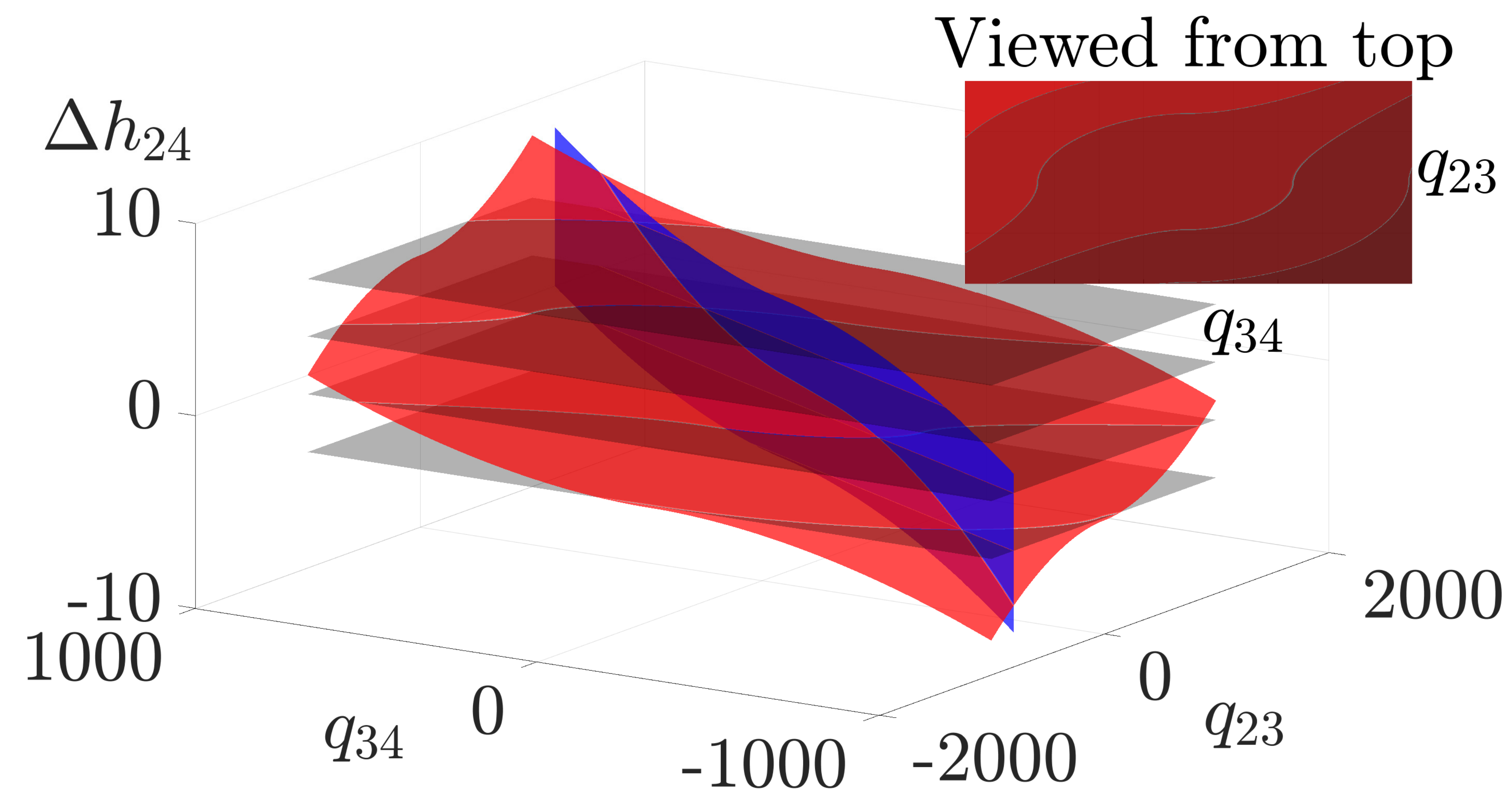}
			\label{fig:3nodes3dv2}
		}
		\vspace{-0.5em}
		\caption{3-node network and  visualization of its equations.}
		\label{fig:3nodesvisual}
	\end{figure}
	The bottleneck of this algorithm is solving a scalable LP/QP successively which should not cause a large computational burden, even if tens of iterations are required to converge. We note that Steps 7 and 8 are used to accelerate the computational times, since the direction of the search is known, and the acceleration parameter $a$ in Step 8 is needed to be adjusted according to the specific WDN. This will be investigated in future work. We finally note that Algorithm~\ref{alg:gp-alg} does not show the windowing process of performing real-time SE, as the algorithm only shows batch state estimation. However, a moving horizon window can be implemented within Algorithm~\ref{alg:gp-alg} thereby allowing real-time state estimation. 
	
	\begin{table}[t]
		\caption{Formulations of 3-node network (time index $k$ is ignored).}
		\def\arraystretch{1.2}
		\begin{tabular}{c|c}
			\hline
			\multicolumn{2}{c}{{\small \textbf{Original formulation}}} \\ \hline
			\multicolumn{2}{c}{\parbox{7.5cm}{
					\begin{align}
					{\large \min}\;\;\;\;& \norm{\Phi_{23}^\mathrm{P}(q_{23}) + \Phi_{34}^\mathrm{P}(q_{34}) - \Delta \widetilde{h}_{24}}  \label{equ:se-classical}  \\
					\mathrm{s.t.}\;\;\;& q_{23} - q_{34} =d_3 \notag
					\end{align}
					\vspace{-1em}
			}}     \\ \hline
			{\small \textbf{SE formulation}}      & {\small \textbf{QP-SE formulation} }         \\ \hline
			\hspace{-8pt}
			\parbox{4cm}{	
				\begingroup
				\setlength\arraycolsep{2pt}
				\begin{align}
				\min \;\;&   { \epsilon^\top}  { \epsilon} \label{equ:3node-se} \\
				\mathrm{s.t.}\;\; 0 &=  \begin{bmatrix} 
				1 & -1
				\end{bmatrix}
				\begin{bmatrix} 
				q_{23} \\
				q_{34} 
				\end{bmatrix} +   d_3   \notag \\
				{\m \Delta \m h^{\mathrm{P}}} &= \begin{bmatrix}
				\Phi_{23}^\mathrm{P}(q_{23}) \\ 
				\Phi_{34}^\mathrm{P}(q_{34}) 
				\end{bmatrix} 
				\notag
				\end{align}
				\endgroup
			}          & 
			\hspace{-8pt}
			\parbox{4cm}{	
				\begingroup
				\setlength\arraycolsep{2pt}
				\begin{align}
				\min \;\;\;\;&   { \epsilon^\top}  { \epsilon}  \label{equ:3node-qp-se} \\
				\mathrm{s.t.}\;\;	0 &=  \begin{bmatrix} 
				1 & -1
				\end{bmatrix}
				\begin{bmatrix} 
				q_{23} \\
				q_{34} 
				\end{bmatrix} +   d_3  \notag \\
				\m \Delta \m h^{\mathrm{P}} &= \begin{bmatrix}
				q_{23} \\ 
				q_{34}
				\end{bmatrix} +  \begin{bmatrix}
				C_{23}^{\mathrm{P}} \\ 
				C_{34}^{\mathrm{P}}
				\end{bmatrix} \notag 
				\end{align}
				\endgroup
			}              \\ \hline \hline
		\end{tabular}%
		\label{tab:3node}
	\end{table}
	
	\section{Case Studies}~\label{sec:Test}
	
	We present two simulation examples to illustrate the applicability of our approach. The first 3-node network is used to illustrate the geometric meaning of proposed method, and then we test the 8-node network to illustrate that our approach can handle looped topology. All numerical tests are simulated using EPANET Matlab Toolkit~\cite{eliades2009epanet} on Ubuntu 16.04.4 LTS with an Intel(R) Xeon(R) CPU E5-1620 v3 @ 3.50 GHz. CVX~\cite{cvx} is used to solve the optimization problem. We set the base $b = 1.001$ when converting the variables. All case studies are performed for $T=1$ time-horizon; the head unit is $\mathrm{ft}$; and the flow unit is $\mathrm{GPM}$. All codes, parameters, and tested networks are available in~\cite{shenwangSE}. 
	
\begin{figure}[h]
	\centering
	\includegraphics[width=\linewidth]{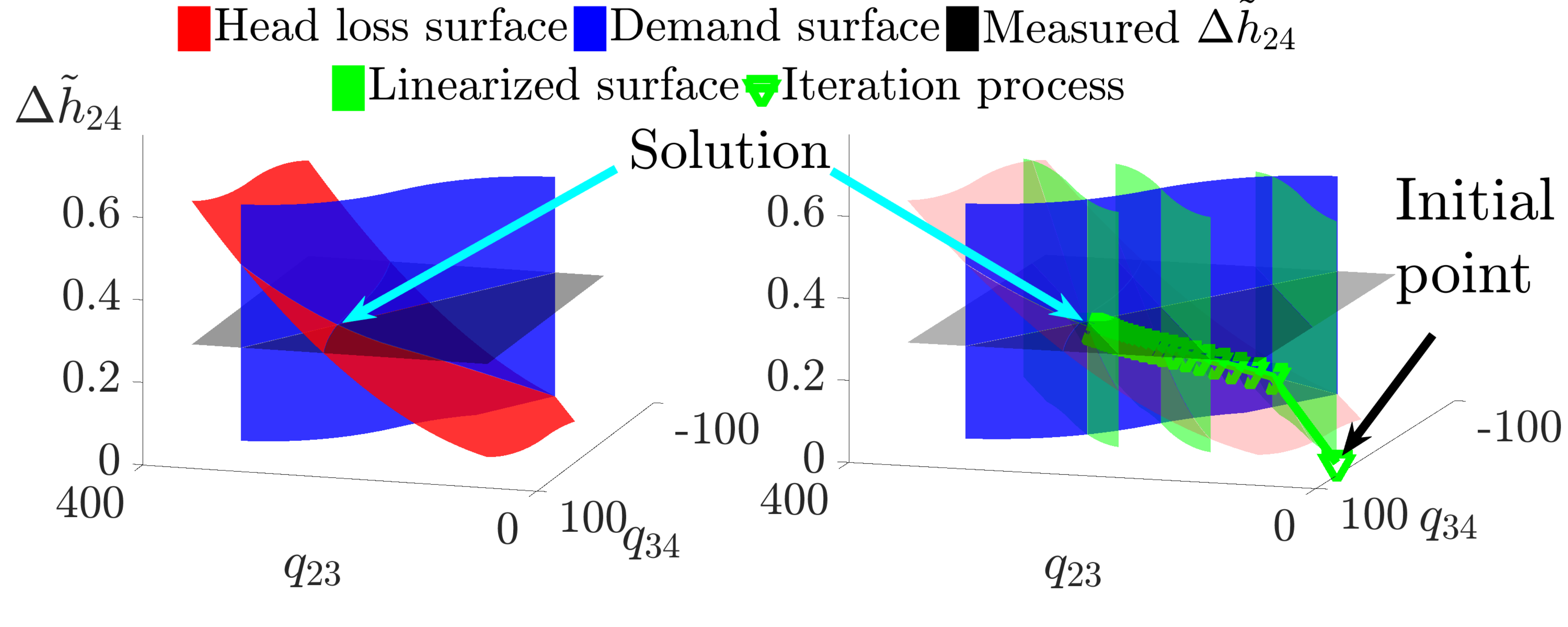}
	\caption{(Left) 3D visualization of~\eqref{equ:se-classical}; (Right) Iteration process of solving scenario 1 (sufficient measurements) for 3-node network.}
	\label{fig:3nodes3dpipeonlyapproximation}
\end{figure}
	\subsection{Three-node Network}
	
	The 3-node network comprised of 3 junctions and 2 pipes is shown in Fig.~\ref{fig:3nodePipes}, and no demand at Junctions 2 and 4. 
	\subsubsection{Sufficient measurements scenario} \label{sec:sufficient}Suppose that we measure head difference $\Delta \widetilde{h}_{24}$  between Junction 2 and 4, and  estimates of the flows  $q_{23}$ and $q_{34}$ are sought. According to~\eqref{equ:classical}, the classical SE  is presented as~\eqref{equ:se-classical} in Tab.~\ref{tab:3node}. In fact, it can be visualized as Fig.~\ref{fig:3nodes3dv2} where the red surface is nonlinear $\Phi_{23}^\mathrm{P}(q_{23})\hspace{-2pt} +\hspace{-2pt} \Phi_{34}^\mathrm{P}(q_{34})$, the blue surface is linear conservation of mass constraint,and the gray surface is measured head difference $\Delta \widetilde{h}_{24}$. The solution is in the intersection of these three surfaces. In order to see the feasible set, we can view 3D plot from top (ignore $\Delta \widetilde{h}_{24}$ dimension). Notice that the feasible set can be viewed as the intersection of red and gray surfaces, and it is highly nonconvex in 2D.

	After conversion, the corresponding QP form of SE~\eqref{equ:3node-se}  is presented as~\eqref{equ:3node-qp-se} in Tab.~\ref{tab:3node}, where $
	\epsilon =\begin{bmatrix}
	1  &1
	\end{bmatrix} \m \Delta \m h^{\mathrm{P}} \hspace{-2pt}-  \hspace{-2pt}
	\Delta \widetilde{h}_{24}$. Iteration process is presented in Fig.~\ref{fig:3nodes3dpipeonlyapproximation}, and intersection of blue and red surface in the left plot are approximated by the intersection of blue surface and multi-green surfaces in the right plot. As we mentioned, this is a new type of linear approximation but not the first order Taylor series of the nonlinear function, we can notice this from the green surfaces.
	\subsubsection{Over-determined measurements scenario} Suppose that we measure two head differences $\Delta \widetilde{h}_{23}$ and  $\Delta \widetilde{h}_{24}$ and that the $\Delta \widetilde{h}_{23}$ is ten times more accurate than  $\Delta \widetilde{h}_{24}$. Therefore, the  weight in the objective function must be updated. Hence, the SE problem can simply be presented as
	\begin{align}
	\min\;& \norm{\Phi_{23}^\mathrm{P}(q_{23})\hspace{-2pt} -\hspace{-2pt}  \Delta \widetilde{h}_{23}}\hspace{-2pt}   + \hspace{-2pt} 0.1\norm{\Phi_{23}^\mathrm{P}(q_{23})\hspace{-2pt}  +\hspace{-2pt}  \Phi_{34}^\mathrm{P}(q_{34}) - \Delta \widetilde{h}_{24}}  \notag \\
	\mathrm{s.t.}\;& q_{23} -  q_{34} =d_3 \label{equ:se-classical2} 
	\end{align}
	With such changes, the objective function in~\eqref{equ:3node-qp-se} becomes
	$ {\m \epsilon^\top} \mathrm{diag}(1,0.1) {\m \epsilon} $
	and the corresponding expression is 
	\begin{align*}
	\epsilon =
	\begin{bmatrix}
	1 & 0\\
	1  &1
	\end{bmatrix} \m \Delta \m h^{\mathrm{P}}-  \begin{bmatrix}
	\Delta \widetilde{h}_{23}\\ 
	\Delta \widetilde{h}_{24}
	\end{bmatrix},
	\end{align*}
	and notice that it is still a QP. In order to prove the effectiveness of our approach, our results is compared with solutions from other solvers. On one hand, the nonlinear optimization problems~\eqref{equ:se-classical} and~\eqref{equ:se-classical2} can be solved optimally via \texttt{fmincon} with \texttt{GlobalSearch} option in Matlab, on the other hand, it can be solved via Algorithm~\ref{alg:gp-alg}. The final results and comparisons are listed in Tab.~\ref{tab:3node-result}. We can see that the proposed algorithm yields similar solutions to \texttt{fmincon} and if measurement $\Delta \widetilde{h}_{23}$ is more reliable than $\Delta \widetilde{h}_{24}$, then final results change accordingly.
	
	\begin{table}[t]
		\scriptsize
		\def\arraystretch{1.3}
		\caption{Results of 3-node and 8-node network.}
		\centering
		\makegapedcells
		\setcellgapes{0.2pt}
		\begin{subtable}[Results for the 3-node network.]{
				\resizebox{\linewidth}{!}{%
					\begin{tabular}{c|c|c|c|c}
						\hline
						& \multicolumn{2}{c|}{\textit{\makecell{Sufficient \\ measurements scenario}}} & \multicolumn{2}{c}{\textit{\makecell{Over-determined \\measurements scenario}}} \\ \hline
						\textit{Variables}& $q_{23}$     & $q_{34}$     & $q_{23}$     & $q_{34}$    \\ \hline
						\textit{\makecell{Fmincon with \\
								GlobalSearch} }   & 238.607      & 38.607       & 234.690      & 34.690      \\ \hline
						\textit{Algorithm~\ref{alg:gp-alg}}  & 238.538      & 38.528       & 235.007      & 35.007      \\ \hline \hline
					\end{tabular}
					\label{tab:3node-result}
				}
			}
		\end{subtable}
		\begin{subtable}[Results for the 8-node network.]{
				\resizebox{\linewidth}{!}{%
					\begin{tabular}{c|c|c|c|c|c}
						\hline
						\multicolumn{2}{c|}{}             & \textcolor{blue}{$h_8^{\mathrm{TK}}$}     & \textcolor{red}{$h_3$ }    & $h_5$     & $q_{46}$   \\ \hline
						\multicolumn{2}{c|}{\textit{True value: }$\m \xi_{\mathrm{EPANET}}$}   & 834.00 & 875.89 & 863.49 & 82.50 \\ \hline
						\multicolumn{2}{c|}{\textit{Measurement: }$ \tilde{\m \xi}$} & \textcolor{blue}{834.60} & \textcolor{red}{875.64}  & ---     &  ---    \\ \hline
						\multirow{2}{*}{\makecell{\textit{Estimation from}\\ \textit{Algorithm~\ref{alg:gp-alg}} $(\m \xi_{\mathrm{SE}})$}}    & \textit{Case 1 }   & \textcolor{blue}{834.56}  & 876.40 & 864.01 & 82.44 \\ 
						& \textit{Case 2}    & 833.89 & \textcolor{red}{875.81}  & 863.36 & 83.81 \\ \hline \hline
					\end{tabular}%
					\label{tab:8noderesult}
				}
			}
		\end{subtable}
	\end{table}
	\subsection{8-node Network}
\begin{figure}[t]
		\centering
		\subfigure[8-node network topology and over-determined measurements scenario (additional measurement $h_3$), blue line is head difference  $\Delta \widetilde{h}_{18}$, and red line is $\Delta \widetilde{h}_{13}$.]{%
			\includegraphics[width=0.7\linewidth]{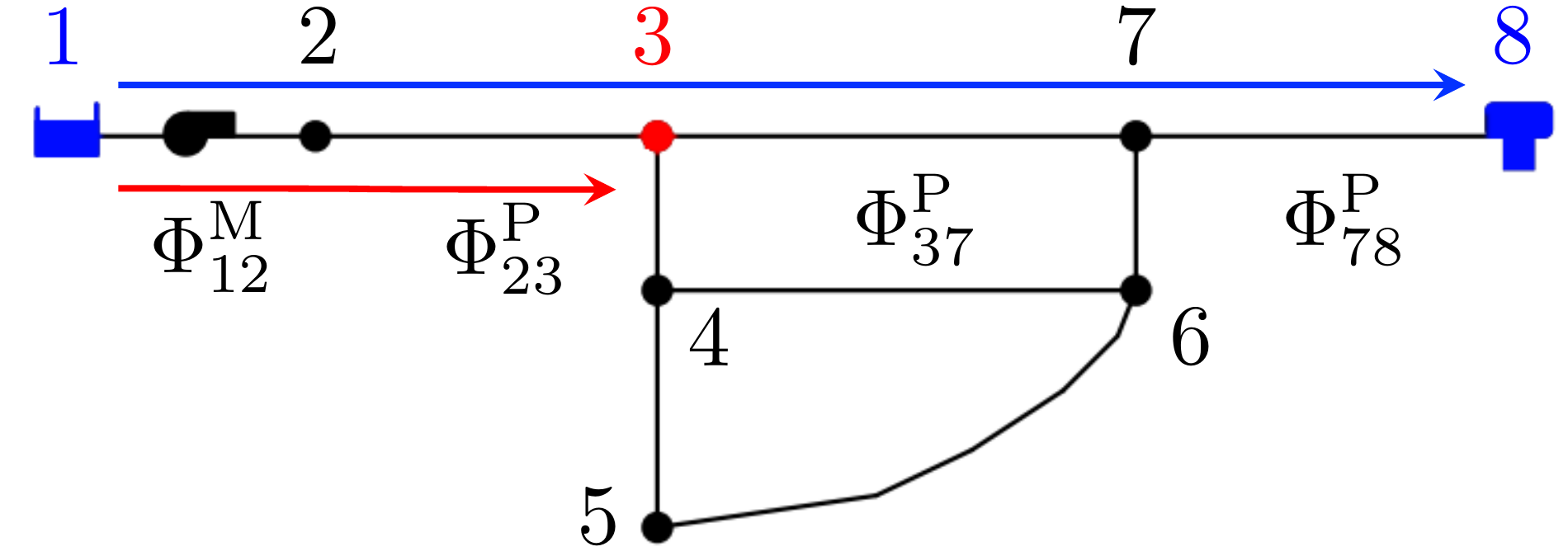}
			\label{fig:8nodenetwork}
		}\\ 
		\subfigure[Error between solution from EPANET and our approach in sufficient measurements scenario.]{%
			\centering
			\includegraphics[width=0.82\linewidth]{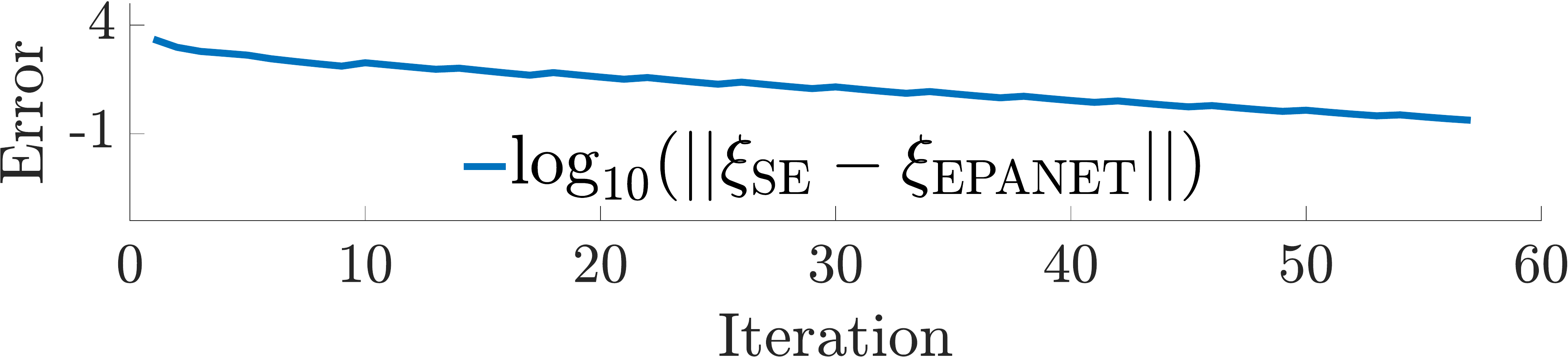}
			\label{fig:errorepanet}
		} 
		\\ 
		\caption{8-node network and its results under multi-scenarios.}
		\label{fig:8nodes}
	\end{figure}

	The  8-node network from EPANET Users Manual~\cite{rossman2000epanet} is a looped network, and labels for various components are all shown in~Fig.~\ref{fig:8nodenetwork}. If the head at reservoir and tank, $h_1^{\mathrm{R}}$ and $h_8^{\mathrm{TK}}$,  are known, then it satisfies the sufficient scenario which means the solution can be determined with just these two measurements. The solution $\m \xi_{\mathrm{SE}}$ solved by Algorithm~\ref{alg:gp-alg} and $\m \xi_{\mathrm{EPANET}}$  from EPANET are given in~Tab.~\ref{tab:8noderesult} when $h_8^{\mathrm{TK}} = 834\ \mathrm{ft}$ and $h_1^{\mathrm{R}}=700\ \mathrm{ft}$, and final error $\norm{\m \xi_{\mathrm{SE}}-\m \xi_{\mathrm{EPANET}} }$ presented in Fig.~\ref{fig:errorepanet} reaches 0.1 which illustrates the effectiveness of proposed method.
	
	As we mentioned in Section~\ref{sec:SE-WDN}, we assume that the measurements of heads at tanks and reservoirs are very accurate. We measure one more head at Junction 3 ($h_3$) thereby defining the over-determined scenario. There are two cases based on which measurement is more trustful. For Case 1, if we postulate that $h_8^{\mathrm{TK}}= 834.60\  \mathrm{ft}$ is more accurate and setting the accuracy matrix as $W = \diag(1,0.1)$, we see the resulting $h_8^{\mathrm{TK}}$ is very close to the measured value in Tab.~\ref{tab:8noderesult}, while $h_3$ is far from its measurement since it is considered less accurate. For Case 2, if $h_3$ is considered more accurate, $h_3$ is close to its measured value $875.64\ \mathrm{ft}$. Besides that, the two estimated variables $h_5$ and $q_{46}$ are shown, and we can see that both are close to the values provided by the EPANET software, but vary slightly between Cases 1 and 2.
	
	\section{Paper's Limitations and Future Work}~\label{sec:Future}
	The paper's limitations include the lack of uncertainty quantification from nodal water demands, leaks in pipes, and pipe roughness parameters. The proposed approach can handle ellipsoidal and cardinal-polyhedral uncertainty by updating the constraints through the successive linear approximation, given historical sets of demands and network parameters. 
	
	To this end, future work will focus on deriving a robust, yet still scalable state estimation routine that addresses uncertainty stemming from the aforementioned sources, in addition to modeling various types of valves in the SE problem, and thoroughly comparing our approach to the state-of-the-art in the literature. 
	
	\section{Acknowledgments} 
	We gratefully acknowledge the constructive comments from
	the conference editor and the reviewers. We also acknowledge
	the financial support from National Science Foundation through
	Grants 1728629 and 1917164.
	\vspace{-0.25cm}
	\bibliographystyle{IEEEtran}
	\bibliography{IEEEabrv,bibfile4}

\end{document}

%% file: preambleT.tex
\usepackage{epsfig,color,amsmath,cite}

\usepackage{tikz}
\usetikzlibrary{arrows.meta,
	backgrounds,
	fit,
	matrix}
\tikzstyle{startstop} = [rectangle, rounded corners, minimum width=1cm, minimum height=0.25cm,text centered, draw=black, fill=red!7]
\tikzstyle{io} = [trapezium, trapezium left angle=75, trapezium right angle=105, text centered, draw=black, fill=blue!7]
\tikzstyle{process} = [rectangle, minimum width=1cm, minimum height=0.25cm, text centered, draw=black, fill=orange!7]
\tikzstyle{decision} = [diamond, text centered, draw=black, fill=green!7]
\tikzstyle{arrow} = [thick,->,>=stealth]

\usepackage{pgfplots,caption}
\pgfplotsset{compat=1.9}

\usepackage{subfigure}

\usepackage{wrapfig}
\usepackage{xcolor}
\usepackage{epstopdf}
\usepackage{amssymb}
\usepackage{url}
\usepackage{mathtools}
\usepackage{enumitem}
\usepackage{multirow}
\usepackage{ifthen}
\usepackage{makecell}
\usepackage{amsmath}
\usepackage{stackrel}
\usepackage{booktabs}
\usepackage[flushleft]{threeparttable}
\usepackage{makecell}
\usepackage{mathabx}
\usepackage[linesnumbered,lined,boxed,commentsnumbered,ruled,longend]{algorithm2e}


\setlist[itemize]{leftmargin=*}

\DeclareMathOperator{\diag}{diag}

\makeatother

\newcommand{\m}{\boldsymbol}
\allowdisplaybreaks[4]
\usepackage[colorlinks = true,
linkcolor = blue,
urlcolor  = blue,
citecolor = blue,
anchorcolor = blue]{hyperref}

\usepackage{graphicx} 
\graphicspath{{Figures/}}
\usepackage{amsthm}
\theoremstyle{definition}

\usepackage[framemethod=TikZ]{mdframed}
\usepackage{bm}
\mdfdefinestyle{MyFrame}{%
	linecolor=black,
	outerlinewidth=1.5pt,
	roundcorner=1.5pt,
	innerrightmargin=5pt,
	innerleftmargin=5pt,}

\usepackage{multicol}
\DeclarePairedDelimiter{\norm}{\lVert}{\rVert}